\documentclass[10pt,journal,compsoc]{IEEEtran}

\ifCLASSOPTIONcompsoc
  \usepackage[nocompress]{cite}
\else
  \usepackage{cite}
\fi
\ifCLASSINFOpdf
\else
\fi
\usepackage{amsmath}
\usepackage{graphicx}
\begin{document}
%
\title{CD-SFA: Stochastic Frontier Analysis Approach to Revenue Modeling in Cloud Data Centers}
%
%
%
%
\title{\LARGE \bf
CD-SFA: Stochastic Frontier Analysis Approach to Revenue Modeling in Large Cloud Data Centers}
\author{~Jyotirmoy Sarkar, Bidisha Goswami,~\IEEEmembership{Member,~IEEE,}
        Snehanshu Saha,~\IEEEmembership{Senior Member,~IEEE,}
        and~Saibal~Kar~\IEEEmembership{}
\IEEEcompsocitemizethanks{\IEEEcompsocthanksitem J. Sarkar is with Wipro GE Healthcare, Bangalore, India.\protect\\
E-mail: jyotirmoy208@gmail.com
\IEEEcompsocthanksitem S.Saha and B.Goswami are with PESIT South Campus, Bangalore, India.
\IEEEcompsocthanksitem S.Kar  is with the Deaprtment of Economics, Calcutta University and Centre for Studies in Social Sciences, Calcutta.}
\thanks{Manuscript received , ; revised , .}}

%
%

\markboth{}%
{Shell \MakeLowercase{\textit{et al.}}: Bare Advanced Demo of IEEEtran.cls for IEEE Computer Society Journals}
%



\IEEEtitleabstractindextext{%
\begin{abstract}
Enterprises are investing heavily in cloud data centers to meet the ever surging business demand. Data Center is a facility, which houses computer systems and associated components, such as telecommunications and storage systems. It generally includes power supply equipment, communication connections and cooling equipment. A large data center can use as much electricity as a small town. In today’s world due to the emergence of data-center based computing services, it has become necessary to examine how the costs associated with data centers evolve over time, mainly in view of efficiency issues. We have presented a quasi form of Cobb-Douglas model, which  addresses revenue and profit issues in running large data centers. The stochastic form has been introduced and explored along with the quasi Cobb-Douglas model to understand the behavior of the model in depth. Harrod neutrality and Solow neutrality are incorporated in the model to identify the technological progress in cloud data centers.This allows us to shed light on the stochastic uncertainty of cloud data center operations. A general approach to optimizing the revenue/cost of data centers using Cobb Douglas Stochastic Frontier Analysis( CD-SFA) is presented. Next, we develop the optimization model for  large data centers. The mathematical basis of CD-SFA has been utilized for cost optimization and profit maximization in data centers. The results are found to be quite useful in view of production reorganization in large data centers around the world.
\end{abstract}

}

\maketitle

\IEEEdisplaynontitleabstractindextext

%
\IEEEpeerreviewmaketitle

\ifCLASSOPTIONcompsoc
\IEEEraisesectionheading{\section{Introduction}\label{sec:introduction}}
\else
\section{Introduction}
\label{sec:introduction}
\fi

%
%
%
%
\IEEEPARstart{D}{ata} center is an integral part of IT organizations for running everyday business operations. A data center is a virtual or physical centralized facility, which is used for storing, managing information and data associated with an organization. Cloud and traditional data centers have a few differences albeit, both are used for storing data. The striking difference between a cloud and a traditional data center is that cloud is an off-premise form of computing that stores data on the Internet, whereas a traditional data center uses on-premise hardware that stores data within an organization's local network[1]. A typical data center is an in-house facility managed by organization's own IT team, whereas cloud is generally managed and maintained by the third party. The shift
from PC-based computing services to server-side computing is driven primarily not only by the improvements in services, such as the ease of management (no configuration or backups needed) and ubiquity of access (a browser
is all you need), but also by the advantages it offers to vendors.
Data center economics allows many application services to run at a low cost per user. For
example, servers may be shared with thousands of active users (and many more inactive ones), resulting in better utilization. Similarly, the computation itself may become cheaper in a shared service (e.g., an email attachment received by multiple users can be stored once rather than
many times). 
For constructing a tier 1 data center, which houses 5 to 10 racks along with Cooling, Power (UPS + Generator), Fire Suppression etc., the company needs to incur a monthly fixed cost between US\$4046 and US\$6100, apart from the subsequent variable maintenance cost. The maintenance cost of the existing infrastructure  varies from \$2100 to \$4500 each month [2]. In view of the substantial costs and economic rent associated with these data centers, it is imperative that dynamic optimization exercises are carried out periodically. The choice of optimum size of the data center (if factors and product prices are exogenous) just by itself, may lead to Pareto improvements globally, given the ever expanding depth and width of IT use.\\  This paper finds the profit maximizing levels of operations in the large data centers. Of course, technological improvements may be a crucial element for such optimization. A specific form of technological improvement, which increases the efficiency of labor, called Harrod-neutral technical progress, has been discussed in the paper. Harrod-neutral technological progress allows producers to make more output by investing less. In other words, since both capital and labor become more productive, the technical progress relaxes the constraints and promotes growth. In the basic Solow model, also discussed and implemented for the first time in Cloud Data Centers, growth occurs only as a result of factor accumulation.

Because the technology has the neoclassical form (diminishing returns to per capita capital), capital accumulation cannot raise per capita income forever. Solow model also proposes that the long run growth is achievable only through  changes in technology. The change in the capital investment is possible from the change in the savings rate. In the short run, growth can be defined by moving towards a steady state, which is created by accumulated factors such as capital and labor force growth. In order to run a data center, skilled and efficient labor force is necessary. In traditional data centers, as much as 40\% of the cost is associated with labor alone, whereas labor costs are 6\% of the total operating cost of a cloud data center. Innovation of new business processes and technological progress may reduce the cost and enhance the knowledge of workers. Maintaining the existing infrastructure along with cooling systems and power back up systems needs huge amount of annual capital investment. Therefore  innovation in server technology, which dissipates less heat or new cooling equipment, power backup generator which absorb less electricity can reduce the capital cost for running a data center. Hence Solow neutrality technological progress, which is capital augmentation is as important as Harrod technological progress.
A production function will be called frontier when it gives the maximum possible output for a given set of inputs. All the production units of a frontier function will be fully efficient.Now, efficiency can be explained in two ways: technical and allocative.The technical efficiency can be further modeled by either deterministic or by stochastic frontier production function. The deterministic frontier model explains the shortfall from the frontier, which is the maximum output by technical inefficiency, whereas the stochastic model includes the random shocks to the frontier function [6].\\
The remainder of the paper is organized as follows: A brief literature survey is offered in section 2. Section 3 discusses the analytical basis and the results of the optimization carried out with real time data from large data centers. We proposed a quasi Cobb-Douglas production function, which has been integrated with Harrod and Solow neutrality of technological progress. We have tried to establish the relevance of Cobb-Douglas function in a data center. Stochastic Gradient Descent is used rigorously to find out the optimized cost associated with a data centers. Section 4 concludes the model with supporting results. All mathematical proofs, computation of technological progress and matlab codes are available in the appendix of the additional file [10]. 
\section{Related Work}
\IEEEPARstart{J}{ames} Hamilton [5] has shown that, quite significantly, power is not the largest cost, if the amortization cost of power, cooling infrastructure for 15 years and new server amortization cost over 3 years are taken into consideration. He concluded that, if the monthly payments for cooling amortization and server amortization are computed using 5\% per annum cost, then server hardware costs
are the largest. \footnote{However, it was also pointed out that costs due to power consumption might rise in the future, while the server hardware costs fall. Overall, it might make power-cost dominate all other items in the cost function.} The interface of Cobb-Douglas production or cost functions, a frequently used 'well-behaved' functional form in economics (see Varian, 1992), finance and related disciplines, and that of the Stochastic Frontier Analysis, does not deal with the present research question quite often. In fact, the use of SFA with Cobb-Douglas production or cost functions is regularly used in the analysis of banking efficiency in different countries (see, Sensarma, 2004; Lozano-Vivas, 1997; Aly, et al., 1990, etc.), or agricultural crop efficiency (viz, Battesse and Coelli, 1992), but rarely to measure the cost efficiency of large data centers. This not only expands the horizon of applications beyond a handful of subjects, but also exemplifies pragmatic use of this interface in industry, directly. Cobb-Douglas function has been widely used in economics and various sectors. 
Empirically speaking,De-Min Wu [3], have shown the exact distribution of the indirect least squares estimator of the coefficients of the Cobb-Douglas production function within the context of a stochastic production model of Marschak-Andrews type. Efstratios Rappos, Stephan and Rudlof have proposed integer programming optimization model of data center for determining the optimal allocation of data components among a network of Cloud data servers in such a way that it minimizes the total costs of additional storage, estimated data retrieval costs and network
delay penalties [4].  Saha et al. proposed an algorithmic/analytical approach to address the issues of optimal utilization of the resources towards a feasible and profitable model in running Cloud Data Centers [11]. The model suggests minimum investments needed to achieve target output.
\section{Analytical Foundation}
IT organizations need to invest heavily in order to set up their own data centers  and at times, it may be cheaper to rent space for this purpose rather than construct a new facility. Clearly, this involves certain trade-offs with regard to different costs of data centers. We intend to optimize the cost structure for data centers with the help of a quasi Cobb-Douglas production function. We will start with a brief description of Cobb-Douglas production function.
\begin{equation}\label{eq1}
Y=P L^\alpha K^\beta
\end{equation}
Where Y= total production output\\
      L=Labor input\\
      K=Capital input\\
      P=Total factor productivity \\
As discussed above, the two widely used mathematical models namely Harrod and Solow neutral progress to predict the technological progress can be integrated with equation (\ref{eq1}) to accommodate time variant technological changes.

\begin{equation}\label{eq2}
Y=P [A_i(t)L]^\alpha K^\beta
\end{equation}

\begin{equation}\label{eq3}
Y=P L^\alpha [B_i(t)K]^\beta
\end{equation}
 Combining equations (\ref{eq2}) and (\ref{eq3}) .
\begin{equation}\label{eq4}
Y=P [A_i(t)L]^\alpha [B_i(t)K]^\beta
\end{equation}
\begin{equation}\label{eq5}
Y=P [AL]^\alpha [BK]^\beta
\end{equation}
We have assumed technological progress A and B as endogenous variables, hence dependent on other parameters related with R\&D. Therefore, Harrod neutral technological progress A and Solow neutral technological progress B may be represented as follows:
\begin{align}\label{eq6}
A=rL^{*\beta_1}\Gamma^{1-\beta_1}
\end{align}
where r is the future discount rate. \\
$L^{*}$ is the labour involved in R\&D related to Harrod technological progress.\\
$\Gamma$ is the capital invested for R\&D 
\begin{align}\label{eq7}
B=rK^{*\alpha_1}\Delta^{1-\beta_1}
\end{align}
where r is the future discount rate as usual. \\
$K^{*}$ is the capital invested in R\&D related to Harrod technological progress whereas $\Delta$ is the labour contribution to R\&D.
\subsection{Revenue Maximization}
Consider an enterprise that has to choose its consumption bundle (I,R) where I, R are infrastructure and recurring costs respectively of a cloud data center. The enterprise wants to maximize its production, subjected to the constraint that the total cost of the bundle does not exceed a particular amount. The company has to keep the budget constraint in mind and keep total spending within this amount. The production maximization is done using Lagrangian Multiplier. The quasi Cobb-Douglas function is formulated as:
\begin{equation}\label{eq8}
f(I,R)=[AR]^\alpha [BI]^\beta
\end{equation}
Let m be the cost of the inputs that should not be exceeded.
\begin{equation}\label{eq9}
w_1AR+w_2BI=m
\end{equation}
\(w_1\) is the unit cost of augmented recurring cost.\\
\(w_2\) is the unit cost of augmented infrastructure cost.
Optimization problem for production maximization is written as:\\
\hspace{1cm} \hspace{1cm}       max \(f(I,R)\) subject to m
The following values of A,B obtained are the values for which the data center has maximum production after satisfying the constraints on the investment.
\begin{equation}\label{eq10}
A=\frac{\alpha m}{w_1 R(\alpha+\beta)}
\end{equation}
\begin{equation}\label{eq11}
B=\frac{\beta m}{w_2 I(\alpha+\beta)}
\end{equation}

Replacing the values of A and B by using equations (\ref{eq6}) and (\ref{eq7}).
\begin{align*}
L^{*}=\frac{m\alpha}{rw_1R\left(\alpha+\beta\right)\Gamma^{1-\beta_{1}}}^{\frac{1}{\beta_1}}\\
K^{*}=\frac{m\beta}{rw_2I\left(\alpha+\beta\right)\Delta^{1-\alpha_1}}^{\frac{1}{\alpha_1}}
\end{align*}
These results are proved in Appendix A, Additional file [10].
\subsection{Cost Minimization}
Consider an enterprise with a target level of output to achieve by investing a minimum amount. The quasi Cobb-
Douglas function is of the form:\\
\(f=[AR]^\alpha[BI]^\beta\)
where \(y_{tar}\) is the target output of the firm that needs to be
achieved and w1,w2 are unit prices of Recurring cost,and infrastructure respectively. Cost minimization problem is formulated as follows:\\

\(min_{A,B} w_1AR+w_2BI\) subject to \(y_{tar}\)

The cost of producing \(y_{tar}\) units in cheapest way is c,
where
\begin{equation}
c=w_1AR+w_2BI
\end{equation}
c can be written as follows:
\begin{equation}
c=\frac{w_1y_{tar}^\frac{1}{\alpha+\beta}}{\frac{\beta w_1}{\alpha w_2I}^\frac{\beta}{\alpha+\beta}}+\frac{w_2y_{tar}^\frac{1}{\alpha+\beta}}{\frac{\alpha w_2}{\beta w_1 R}^\frac{\alpha}{\alpha+\beta}}
\end{equation}
The details of the above results have been elaborated in Appendix B, Additional file [10].
\subsection{Profit Maximization}
Consider an enterprise that needs to maximize its profit.
The Profit function is:
\begin{equation}
F=f(A,B)-w_1AR-w_2BI
\end{equation}
Profit maximization is achieved when :\\
\(\frac{\partial f}{\partial A}=w_1R \hspace{.3cm} and \hspace{.3cm} \frac{\partial f}{\partial B}=w_2I\)\\
The values are obtained after the the calculations as below:\\
\begin{equation}
A=\frac{w_1^\frac{1}{\alpha+\beta-1}}{R \alpha^\frac{1-\beta}{\alpha+\beta-1}\left(\frac{w_1\beta}{w_2}\right)^\frac{\beta}{\alpha+\beta-1}}
\end{equation}
\begin{equation}
B=\frac{w_2^\frac{1}{\alpha+\beta-1}}{I \beta^\frac{1-\alpha}{\alpha+\beta-1}\left(\frac{w_2\alpha}{w_1}\right)^\frac{\alpha}{\alpha+\beta-1}}
\end{equation}
The above results are proved in Appendix C, Additional file [10]. Substituting the values in Equation (\ref{eq5}) we obtain
\begin{equation}
Y=P\frac{w_1^\frac{1}{\alpha+\beta-1}}{ \alpha^\frac{1-\beta}{\alpha+\beta-1}\left(\frac{w_1\beta}{w_2}\right)^\frac{\beta}{\alpha+\beta-1}}^\alpha \frac{w_2^\frac{1}{\alpha+\beta-1}}{ \beta^\frac{1-\alpha}{\alpha+\beta-1}\left(\frac{w_2\alpha}{w_1}\right)^\frac{\alpha}{\alpha+\beta-1}}^\beta
\end{equation}

Replacing the values of A and B by equations (\ref{eq6}) and (\ref{eq7}), the following results are derived.\\
\begin{align*}
L^{*}=\frac{w_1^\frac{1}{\alpha+\beta-1}}{rR \alpha^\frac{1-\beta}{\alpha+\beta-1}\left(\frac{w_1\beta}{w_2}\right)^\frac{\beta}{\alpha+\beta-1}\Gamma^{1-\beta_1}}^{\frac{1}{\beta_1}}\\
K^{*}=\frac{w_2^\frac{1}{\alpha+\beta-1}}{rI \beta^\frac{1-\alpha}{\alpha+\beta-1}\left(\frac{w_2\alpha}{w_1}\right)^\frac{\alpha}{\alpha+\beta-1}\Delta^{1-\alpha_1}}^{\frac{1}{\alpha_1}}
\end{align*}
We conclue that the output revenue, in case of profit maximization is independent of labor and capital input.
\subsection{Stochastic frontier}
The production frontier can be written as:
\begin{align*}
y=f(K,L)TE
\end{align*}
where TE is the technical inefficiency,  the ratio of observed output to maximum possible output. If TE=1, the organization achieves maximum output. This production frontier is deterministic as the entire deviation from maximum feasible output is attributed to technical inefficiency. It does not consider random shocks, which is not beyond control of production function. To address the random shocks, the production frontier function can be redefined as below:
\begin{align*}
y=f(K,L)TE exp(v)
\end{align*}
where v is the stochastic variable which defines the shocks, uncertainty, luck etc.
Let us consider the linear logarithmic form of stochastic frontier production function.
\begin{align}
\ln y=K+\alpha \ln S+\beta \ln I+v-u
\end{align}
where y =output\\
S=Server cost of data center\\
I=infrastructure cost\\
v=random shocks\\
u=technical inefficiency
\begin{equation}
\alpha+\beta=n
\end{equation}
CRS: n=1* Constant returns to scale*\\
IRS: \( n>1\)* Increasing returns to scale*\\
DRS: \(n<1\)* Decreasing returns to scale*\\
By solving these two equations, the following values of elasticity can be derived.
\begin{equation}
\alpha=\frac{\ln y-K-\ln I-v+u}{\ln\frac{S}{I}}
\end{equation}
\begin{equation}
\beta=\frac{\ln y-K-\ln S-v+u}{\ln\frac{I}{S}}
\end{equation}
The detailed proof is contained in Appendix D, Additional file [10].
\section{Result \& Discussion}
\begin{figure}[h]\label{Fig 1}
\centering
\includegraphics[scale = 0.7]{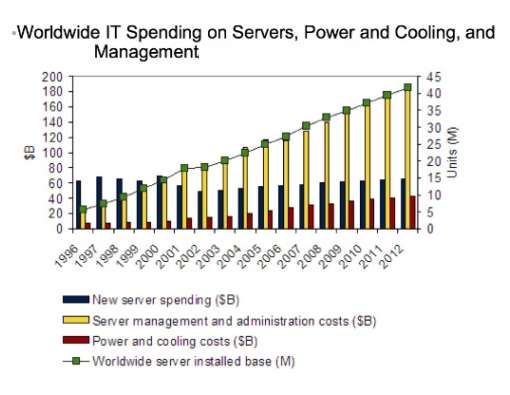}
\caption{World Wide IT spending}
\end{figure}

As we have mentioned earlier that server and power/cooling cost are the major cost segments among all the cost segments related to data center. These two cost segments will be considered for optimal cost and maximum revenue calculations, though cobb-douglas function can accommodate any number of inputs. It is also feasible to aggregate two or more cost segments into two broad categories and use these in the proposed cost model.
The data associated with data center costs from various sources have been accumulated and Stochastic Gradient Ascent/Descent algorithm is applied on varying elasticities/ unit cost to find the optimal solution for revenue and cost of data center. All simulation results have been generated by a computer system using Matlab. The approximate data from the \ref{Fig 1} for two types of costs, namely server management/administrative cost and power/cooling cost are captured.

Let us now explore Stochastic Gradient Descent method, which has been used thoroughly on real time data set to evaluate optimum revenue and cost along with profit. Stochastic Gradient Descent is a well known method and used in many different fields to achieve optimal value.
\subsection{SGD for Cost minimization}
Here, we have adopted a different approach to find out the values of cost optimization. Generally a cost function is represented as a linear function, which is nothing but a summation of different cost segments involved in optimization. Cobb-Douglas production function has been used to represent cost function. Hence the cost function can be rewritten as follows:
\begin{align*}
c=L^\alpha K^\beta
\end{align*}
Stochastic Gradient descent is used to determine the minimal cost of a data center, when the input server and infrastructure cost are known. In simple terms, we tried to identify the elasticity of server and infrastructure, for which minimum cost is attained under certain constraints. \\
\underline{\textbf{SGD algorithm}}
\begin{itemize}
\item Choose an initial vector of parameters \(\alpha\), \(\beta\) and randomly select learning rate $\delta$
\item \(\frac{\partial c}{\partial \alpha}=\alpha L^{\alpha-1}K^\beta\)
\item \(\frac{\partial c}{\partial \beta}=\beta L^{\beta-1}K^\alpha\)
\item Repeat\\
\textit{Rather than calculating the gradient once, which happens in conventional gradient algorithm, here for each iteration the gradient being recalculated and subtracted from the updated $\alpha$ and $\beta$ }
\item \(\alpha_{n+1}\leftarrow \alpha_{n}-\delta \frac{\partial c}{\partial \alpha}\)
\item \(\beta_{n+1}\leftarrow \beta_{n}-\delta \frac{\partial c}{\partial \beta}\)
\item \(\alpha_{n}\leftarrow \alpha_{n+1}\)
\item \(\beta_{n}\leftarrow \beta_{n+1}\)\\
\textit{The iteration will continue till both the values are greater than 0}
\item until $\left( \left(\alpha_{n+1}> 0) || (\beta_{n+1}> 0\right)\right)$\\
\textit{Stop when the convergence conditions are met}
\item Calculate the cost  by putting \(\alpha, \beta\) in the cost function.
\end{itemize}
\begin{table}[h]
\caption{Sample Table of Simulation output for cost minimization using CD form, Detailed table in Additional file, Table 1 [10]}\label{table:1}
\centering
\begin{tabular}{c c c c c c} 
\hline 
Year&New Server&P\& C&\(\alpha\)&\ \(\beta\)&Min. Cost \\ [0.5ex]
\hline
 
1997 & 65 & 5 & 0.4615  & $6.1674*10^-05$ & 6.8672 \\ 
2002 & 45 & 15 & 0.3338  &  0.0019 & 3.5813\\ 
2009 & 58 & 30 & 0.2416  & $5.1719*10^-04$  & 2.6715\\ 
2012 & 60 & 40 & 0.1670  & $7.6964e-04$   & 1.9872 \\ \hline
\end{tabular}
\end{table}
The approximate cost of New server and Power \& Cooling  represented in Table 1  are collected from the figure 1, whereas $\alpha,\beta$ represent the elasticity of new server and power \& cooling respectively. If the minimum cost table is observed carefully, it is noticed that the minimum cost is gradually decreasing as the year passes with rising investment in power and cooling, whereas the new server cost is almost stable from the year 1996 to 2012. Stochastic Gradient Descent performs quite well when the difference between the cost segments are relatively low. As the difference between new server and power\& cooling cost segments is the lowest in the year of 2010, the minimum cost is attained in that particular year.
\begin{table}[h]
\caption{Sample Table of Simulation output for cost minimization using Linear cost function, Detailed table in Additional file, Table 2 [10]}\label{table:1}
\centering
\begin{tabular}{c c c c c c} 
\hline 
Year&New Server&P\& C&\(w_1\)&\ \(w_2\)&Min. Cost \\ [0.5ex]
\hline
1997 & 65 & 5 & 0.0150  & 0.6550 & 4.25\\ 
2002 & 45 & 15 & 0.0150    &  0.5050 & 8.25\\ 
2009 & 58 & 30 & 0.0200      & 0.4000 & 13.16\\ 
2012 & 60 & 40 & $5.5*10^-17$   & 0.3000  & 12 \\ \hline
\end{tabular}
\end{table}
\begin{figure}[!h]\label{Fig 2}
\centering
\includegraphics[scale = 0.35]{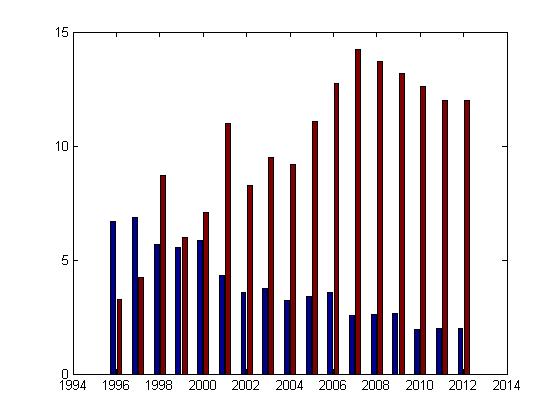}
\caption{Cost Comparison: CD-SFA(Blue) vs Linear Cost(Red) function}
\end{figure}
Fig 2 displays the comparison between two form of cost functions.The blue colored bar represents the minimum cost achieved by CD inspired cost function, whereas the other color bar points to the minimum cost attained by linear cost function. The minimum cost decreased in case of CD inspired cost function as the cost segments ncreased after each consecutive year. In contrast, the minimum cost increased as long as the different cost segments increase in case of linear form of cost function. 
\subsection{SGA for Revenue maximization}
As the objective of any organization is to maximize the revenue, we have employed Stochastic Gradient Ascent algorithm to calculate optimum revenue.
In such scenario, where IT organizations are under constant pressure of budgetary constraints, the optimal elasticity are computed using Stochastic Gradient Ascent algorithm.\\
\underline{\textbf{SGA algorithm} }
\begin{itemize}
\item Choose an initial vector of parameters \(\alpha\), \(\beta\) and randomly select learning rate $\delta$
\item \(\frac{\partial y}{\partial \alpha}=\alpha L^{\alpha-1}K^\beta\)
\item \(\frac{\partial y}{\partial \beta}=\beta L^{\beta-1}K^\alpha\)
\item Repeat\\
\textit{Rather than calculating the gradient once, which happens in conventional gradient algorithm, here for each iteration the gradient being recalculated  and added to the updated $\alpha$, $\beta$}
\item \(\alpha_{n+1}\leftarrow \alpha_{n}+\delta \frac{\partial y}{\partial \alpha}\)
\item \(\beta_{n+1}\leftarrow \beta_{n}+\delta \frac{\partial y}{\partial \beta}\)
\item \(\alpha_{n}\leftarrow \alpha_{n+1}\)
\item \(\beta_{n}\leftarrow \beta_{n+1}\)\\
\textit{The iteration will continue till the sum of $\alpha$ and $\beta$ is less than 1.8}
\item until $\left( \left(\alpha_{n+1}> 0\right) || \left(\beta_{n+1}> 0\right) || \left(\alpha+\beta< 1.8\right)\right)$\\
\textit{Stop when the convergence conditions are met}
\item Calculate the cost  by putting \(\alpha, \beta\) in the revenue function.
\end{itemize}
\begin{table}[h]

\caption{Sample Table of Simulation output for Revenue maximization, Detailed table in Additional file, Table 3 [10]}\label{table:3}
\centering
\begin{tabular}{c c c c c c } 
\hline 
Year&New Server&P \& C&\(\alpha\)&\ \(\beta\)&Max. Rev(CD) \\ [0.5ex]
\hline
1997 & 65 & 5 & 0.5312  & 1.2676 & 70.63 \\ 
2002 & 45 & 15 & 0.6151    &  1.1835 & 256.32\\ 
2009 & 58 & 30 & 0.6612      & 1.1362 & 698.68\\ 
2012 & 60 & 40 & 0.693 & 1.1052 & 1006.59 \\ \hline
\end{tabular}
\end{table}
Similar to the previous section, \ref{Fig 1} data is represented in Table 3 along with elasticity and maximum revenue. After applying the Stochastic Gradient Ascent algorithm, the retrieved maximum revenue is displayed in the last column, where as \(\alpha, \beta \) represent the elasticity values for which the maximum revenue has been attained. There is no significant change of new server cost throughout the years, hence power \& cooling is found to be contributing more towards revenue as power \& cooling cost saw a jump of almost 8 times from 1996 to 2012. 
\subsection{Profit Maximization}
Organizations are focusing more to attain maximum revenue by investing less amount in data center. We have seen in earlier subsections, that optimum revenue and cost have been computed using Stochastic Gradient Descent/Ascent algorithm. But  profit computation is done by subtracting minimum cost from optimal revenue. Profit has risen every year but significant jump in profit is observed in the year 2007 and 2010. The optimal profit surges ahead almost 15 times since 2006.
\begin{table}[h]
\caption{Sample Table of Simulation output for Profit maximization, Detailed table in Additional file, Table 4 [10]}\label{table:4}
\centering
\begin{tabular}{c c c c c} 
\hline 
Year&New Server&P\& C&Max. Profit(CD)& Max Profit(Linear)\\ [0.5ex]
\hline
1997 & 65 & 5 & 64.9679 & 66.38 \\ 
2002 & 45 & 15 & 252.5919 &248.07\\ 
2009 & 58 & 30 & 696.0085   &685.52\\ 
2012 & 60 & 40 & 1004.6028 &994.59\\ \hline
\end{tabular}
\end{table}
To facilitate better understanding of the behavior of the stochastic gradient ascent, we plotted two figures which are insightful in the context revenue optimization. Revenue optimization in the years 2012, 2002, 2009 and 1997 are displayed in the first figure, where x-axis, y-axis and z-axis represent $\alpha, \beta$  and revenue respectively.
\begin{figure}[!h]
\centering
\includegraphics[scale = 0.35]{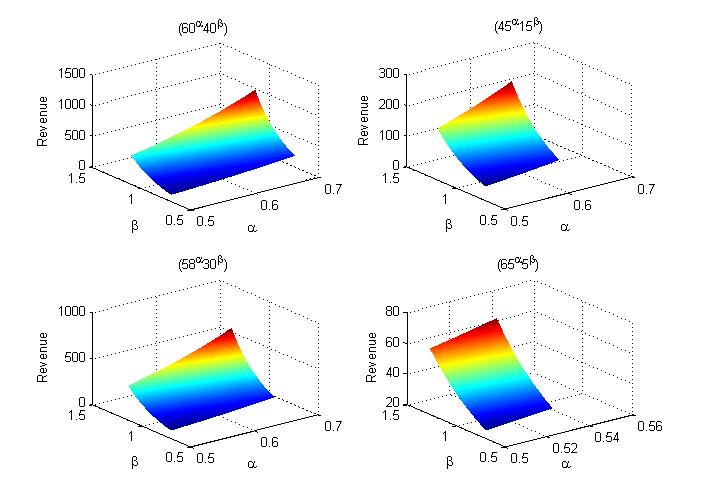}
\caption{Modeled CD-SFA Revenue: Years 2012,2002,2009,1997;}
\label{fig:2}
\end{figure}
Revenue in the years 1999, 2003, 2006 and 2010 are plotted in Figure 4 and is found in\textrm{Additional file, Fig 1 [10]}.\\

It is observed from the figures that revenue rises sharply along with the rise of elasticity ($\alpha \beta$) and no curvature violation is visible. The datapoints which are used in figures are produced by the Stochastic Gradient Ascent algorithm while converging from initial random elasticity values to optimal elasticity values indicating the point where the maximum revenue has been attained. The result produced by the stochastic gradient ascent algorithm has been found to have supporting evidence. For example, the Table \ref{table:3} in the year of 2010 shows that the maximum revenue has been achieved where $\alpha $ is .69 and $\beta$ is 1.1. This fact is also established by the Fig \ref{fig:2}. where  revenue graph touches the peak in the region where $\alpha$ is close to 0.7 and $\beta$ is in the vicinity of 1.1. 
\begin{figure}[!h]
\centering
\includegraphics[scale = 0.35]{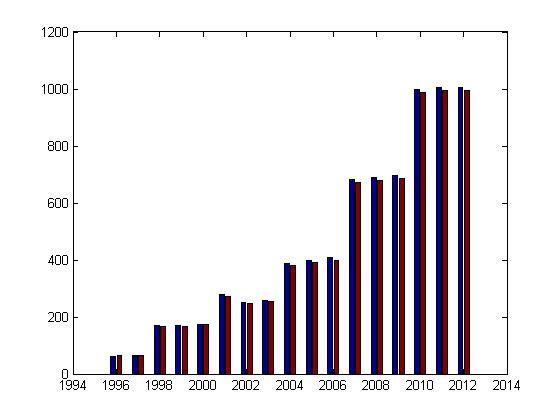}
\caption{Profit Comparison: CD-SFA(Blue) vs Linear Cost(Red) function}
\label{fig:5}
\end{figure}

\begin{figure}[!h]
\centering
\includegraphics[scale = 0.35]{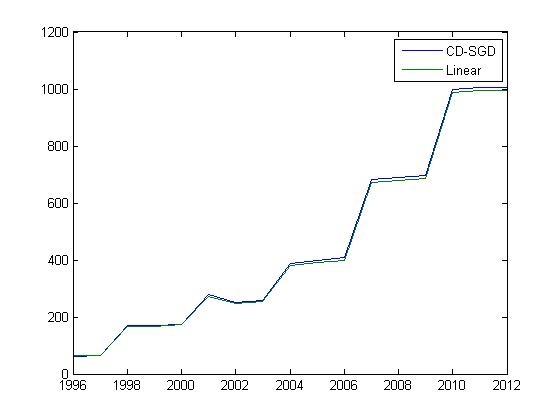}
\caption{Profit comparison CD-SFA vs Linear Year wise }
\label{profitLinear}
\end{figure}
The comparison between maximum profit retrieved in each year by using linear cost and Cobb-Douglas function is plotted in Fig \ref{fig:5}. There is a sharp increase in the profit from 1996 to 2012 except in the year of 2002, where profit has decreased from its previous year. Blue bars point to the maximum profit attained in each year by cost function which is represented by Cobb-Douglas and the other colored bar represents the maximum profit achieved by linear cost function. The figure reflects that the profit achieved by using CD inspired cost function is comparatively higher than the profit attained by using linear cost function. Hence using the nonlinear form of cost function, it is possible to reach maximum profit by controlling the elasticity.
\section{Concentration of firms in Data Center Field}
Profit is found to be increasing with the rising investment in new server and power \& cooling cost. Hence it is required to investigate the level of concentration of firms in the field of Data center. Big firms such Amazone, Google, Microsoft have constructed massive computing infrastructure to support their websites and services. It is imperative to know if  these giants are facing stiff competition. We will use Hirschman-Herfindahl Index (HHI) to know the  competition or concentration of firms. The Herfindahl-Hirschman index (HHI) is a widely used technique to measure the market concentration and it is calculated by squaring the market share of each firm competing in a market, and then summing the resulting numbers. The HHI number can vary from close to 0 to 10,000. The HHI is expressed as:
\begin{equation}
\begin{split}
HHI = s_1^2 + s_2^2 + s_3^2 + ... + s_n^2 
\end{split}
\end{equation}
where \(s_n\) is the market share of the nth firm.\\

High HHI means a few firms control the business. Thus, new cost outlay raises revenue and that increases profit. If there is tough competition in a line of business, the HHI value will be less.
Let us discuss about the market competition in asia pecific region.
The region has generated just over USD 20 billion in data center infrastructure revenues for the world’s leading technology vendors and the market having grown by 23\% from the previous year, according to data from Synergy Research Group [8].\\
\textrm{Fig 6, Data Center Market share in 2011 for Asia Pacific region, is in Additional file, Fig 3 [10] }
\begin{align*}
HHI=21^2+19^2+11^2+8^2+8^2+4^2+4^2+25^2=1708\\
\end{align*}
\textrm{Fig 7, Infrastructure-as-a-Service Market share in 2015 first half, is contained in the Additional file, Fig 2}.

The U.S. Department of Justice considers a market with an HHI of less than 1,000 to be a competitive marketplace; a result of 1,000-1,800 to be a moderately concentrated marketplace; and a result of 1,800 or greater to be a highly concentrated marketplace[7]. In apac region, the concentration is moderate and inclined towards highly concentrated market place. Next we will try to find out the firm concentration in IaaS market place. The IaaS market share data has been collected from a article of businessinsider website [9]. The HHI for IaaS market share is given below\\
\begin{align*}
\begin{split}
HHI&=27.2^2+16.6^2+11.8^2+3.6^2+2.7^2+2.4^2+35.9^2\\
&=2456.34\\
\end{split}
\end{align*}
If we exclude others from HHI calculation, it becomes 1167.53. Still the market can not be considered as competitive. Few firms are controlling the major share of the Infrastructure as a service market place. 
\section{Prediction}
Let the assumed linear form of Cobb-Douglas y = K′ + \(\alpha\) log(S) + \(\beta\) log(P). Consider a set of data points.

\[\begin{array}{ccccccc}
lny_1 & = & K' &{}+{} & \alpha S'_1  & {}+{} & \beta P'_1  \\
 \vdots & & \vdots & & \vdots & & \vdots \\
lny_N & = & K' &{} +{} & \alpha S'_N & {}+{} & \beta P'_N  \\
\end{array}\]

where 
\begin{equation}
S'_i = log(S'_i) \nonumber
\end{equation}
\begin{equation}
P'_i = log(P'_i) \nonumber
\end{equation}

If N \textgreater 3(number of parameters in equation), It is an over-determined system. One possibility is a least squares solution. Additionally if there are constraints on the variables (the parameters to be solved for), this can be posed as a constrained optimization problem. These two cases are discussed below.
\begin{enumerate}
\item No constraints : An ordinary least squares solution is in the form y = Ax where,

$$ x =
 {\begin{bmatrix}
    K' & \alpha & \beta
  \end{bmatrix}}^T
 $$

\begin{equation}
y = \begin{bmatrix}
    y_1\\ . \\ . \\ y_N
  \end{bmatrix}
\end{equation}

and

\begin{equation}
A = \begin{bmatrix}
    1 & S'_1 & P'_1\\ 
      & ...  &     \\ 
    1 & S'_N & P'_N
  \end{bmatrix}
\end{equation}

The least squares solution for x is the solution that minimizes 
\begin{equation}
(y - Ax)^T (y - Ax) \nonumber 
\end{equation} 
It is well known that the least squares solution  is the solution to the system 
\begin{equation}
A^T y= A^TAx \nonumber 
\end{equation}
i.e.
\begin{equation}
x = (A^TA)^{-1} A^T y \nonumber 
\end{equation}

In Matlab the least squares solution to the overdetermined system y = Ax can be obtained by
x = A $\backslash$ y. The following is the result obtained for the elasticity values after performing the least square fitting: 
 
 \begin{table}[!htpb]
\centering
 \begin{tabular}{|c|c|}
\hline 
& Value \\
\hline
\(\alpha\) & 0.2985 \\
\hline
\(\beta\) & 1.3253 \\
\hline
\end{tabular}
\caption{Least square test results}
\end{table}

\item
Constraints on parameters : This results in a constrained optimization problem. The objective function to be minimized (maximized) is still the same namely
\begin{equation}
(y - Ax)^T (y - Ax) \nonumber 
\end{equation} 

This is a quadratic form in x. If the constraints are linear in x, then the resulting constrained optimization
problem is a Quadratic Program (QP). A standard form of a QP is :
\begin{equation}
\text{min  } x^THx +f^Tx
\end{equation}

s.t.

\begin{equation}
Cx \le b \text{   Inequality Constraint}\nonumber
\end{equation}
\begin{equation}
C_{eq}x = b_{eq} \text{   Equality Constraint}\nonumber
\end{equation}

Suppose the constraints are that \(\alpha\)  and \(\beta\) are \textgreater  0 and \(\alpha\)  + \(\beta\)  $\ge$  1. The quadratic program can be written as (neglecting the constant term \(y^T\)y ).

\begin{equation}
\text{min   } x^T(A^TA)x - 2y^TAx
\end{equation}
s.t.
\begin{equation}
 \alpha > 0 \nonumber
\end{equation} 
\begin{equation}
 \beta > 0 \nonumber
\end{equation} 
\begin{equation}
 \alpha + \beta \le 0 \nonumber
\end{equation}

In standard form as given in (26), the objective function can be written as :

\begin{equation}
x^THx + f^Tx
\end{equation}

where

\begin{equation}
H = A^TA \text{  and  } f=-2A^Ty \nonumber
\end{equation}

The inequality constraints can be specified as : 

$$ C =
 \begin{bmatrix}
    0 & -1 & 0\\
    0 &  0 & -1\\
    0 &  1 &  1 
  \end{bmatrix}
 $$

and

$$ b =
 \begin{bmatrix}
    0 \\
    0 \\
    1 
  \end{bmatrix}
 $$
 
 In Matlab, quadratic program can be solved using the function quadprog.
 
The below results were obtained on conducting Quadratic Programming.

\begin{table}[!htpb]
\centering
 \begin{tabular}{|c|c|c|c|}
\hline
 &Value \\
 \hline
 K & 0.1220 \\
 \hline
 \(\alpha\) & 0.4646  \\
 \hline
 \(\beta\) & 1.3354 \\
\hline
\end{tabular}
\caption{Quadratic Programming results}
\end{table}
\item Linear Regression: It is a powerful statistical tool to modeling a relationship between a dependent variable and one or more independent variables. In other words, it tries to predict the dependent variable based on the value of independent variables. One explanatory variable entails simple linear regression. The main focus of linear regression is to fit a single line through a scatter plot. We pose profit as dependent variable and new server and power \& cooling cost as independent variables. The objective is to predict the value of profit by using server and power \& cooling cost.
The linear regression model is as follows
\begin{equation}
y = -375.07 + 4.4871x_1 + 27.409x_2
\end{equation}
Here, y=profit\\
   \( x_1\)=new server cost\\
   \( x_2\)=power \& cooling cost\\
 \(R^2\) value is 0.998\\
  After considering the linear form of Cobb-Douglas, the following linear model is achieved. 
 \begin{equation}
y = 0.82142 + 0.29848 x_1 + 1.3253x_2
\end{equation}
Here, y=profit\\
   \( x_1\)=logarithmic form of new server cost\\
   \( x_2\)=logarithmic form of power \& cooling cost\\
 \(R^2\) value is found to be 0.999 which is a reasonably good fit.\\
 \end{enumerate}

\section{Conclusion}
The paper proposed a model based on quasi cobb-douglas function, focuses on  quantifying the boundary of solow and Harrod neutral technological progress, where optimal revenue, profit and cost occurs. We have endeavored to address the stochastic nature of the production function which is nothing but uncertainty i.e. shocks associated with data center. The deviation from frontier which is the maximum probable output from a data center is explained by model inefficiency, a stochastic variable. Stochastic Gradient Descent and Ascent algorithms have been elaborated and applied on the data set to ensure quick convergence of the elasticity values. Stochastic Gradient Descent is used to estimate maximum revenue, whereas optimal cost is calculated using Stochastic Gradient Ascent. Cost function has been represented as Cobb-Douglas production function rather than a linear function. Graphical representations of the revenue optimization justify our findings from Stochastic Gradient Ascent algorithm.  We have tried to find out the optimal elasticities for which optimal revenue, cost and profit are calculated. But the optimal values of elasticity, derived by Stochastic Gradient Descent/ Ascent, may  fluctuate a little based on the initial assumption of the elasticity and the constraints applied on the model during convergence towards the optimal value. One of the disadvantages of Cobb-Douglas production function is that it is not able to predict the technological progress. We tried to address this shortcoming by accommodating Solow and Harrod technological progress along with the Cobb-Douglas function. Though we have considered  technological progress as constant due to the simplification of calculation and lack of data in Result \& Discussion section. We may conclude that an organization can achieve its target revenue expectation by  controlling the elasticity values. Though the Cobb-Douglas production function can accommodate any number of inputs, but the authors have considered two cost segments associated with data center. We have applied the model on dataset collected from various sources to establish the efficiency. It is clear from our analysis that increasing investment in data center will result in increasing profit. This particular fact is reestablished by HHI index. The HHI is high implying very few organizations control the entire business and competition is less. In such circumstances more investment in server, which is nothing but infrastructure cost, will definitely contribute towards revenue and profit rise. The proposed model is an emphatic testimony of how Large data centers operate and sustain in the long run.


%

\ifCLASSOPTIONcompsoc
  
\fi

\ifCLASSOPTIONcaptionsoff
  \newpage
\fi


\begin{thebibliography}{1}

  \bibitem{}http://www.businessnewsdaily.com/4982-cloud-vs-data-center.html; as accessed on 18/9/2016
  \bibitem{}http://ongoingoperations.com/data-center-pricing-credit-unions/   dated 7/6/2016
   


\bibitem{}Wu D-M (1975) Estimation of the Cobb-Douglas Production
Econometrica 43(4). doi:10.2307/1913082
\bibitem{}Rappos E, Robert S, Riedi RH (2013); A Cloud Data Center Optimization
Approach Using Dynamic Data Interchanges. IEEE 2nd International
Conference on Cloud Networking (CloudNet)
\bibitem{} Hamilton J; (2008) Cost of Power in Large-Scale Data Centers [WWW 1190
document]. http://perspectives.mvdirona.com/2008/11/cost-of-power-
1191in-large-scale-data-centers/; as accessed on 9/3/2015 
\bibitem{}STOCHASTIC FRONTIER MODELS, Camilla Mastromarco, University of Salento, Department of Economics and Mathematics-Statistics
\bibitem{} http://www.investopedia.com/terms/h/hhi.asp; as accessed on
22/1/2016
\bibitem{} https://www.srgresearch.com/articles/hp-ibm-and-dell-leadburgeoning-apac-data-center-infrastructure-market; as accessed on 20/1/2016
\bibitem{} http://www.businessinsider.in/The-cloud-wars-explained-Whynobody-can-catch-up-with-Amazon/articleshow/49706488.cms; as accessed on 10/1/2016
\bibitem{} Additional File, https://www.overleaf.com/read/dmyhyfyjgrry (Read Only)
\bibitem{}Saha, S., Sarkar, J., Dwivedi, A. et al.; A novel revenue optimization model to address the operation and maintenance cost of a data center, J Cloud Computing (2016) 5: 1.; 1-23 doi:10.1186/s13677-015-0050-8

\end{thebibliography}
\end{document}